\documentclass[journal,twoside,final]{IEEEtran}
\usepackage{graphicx}

\ifCLASSINFOpdf
\else
 \fi
\usepackage[cmex10]{amsmath}
\usepackage{cite,algorithmic}
\usepackage{amsfonts,amssymb,array,url,setspace}

\newtheorem{corollary}{Corollary}
\newtheorem{proposition}{Proposition}
\newtheorem{remark}{Remark}

\newcommand{\thres}{x}

\begin{document}

\title{Two-Way Relaying under the Presence of Relay Transceiver Hardware Impairments}
\author{Michail~Matthaiou,~\IEEEmembership{Member,~IEEE,}
        Agisilaos~Papadogiannis,~\IEEEmembership{Member,~IEEE,}\\
        Emil Bj\"ornson,~\IEEEmembership{Member,~IEEE,}
				and M\'erouane Debbah,~\IEEEmembership{Senior~Member,~IEEE} \vspace{-5mm}
\thanks{\copyright 2013 IEEE. Personal use of this material is permitted. Permission from
IEEE must be obtained for all other uses, in any current or future media,
including reprinting/republishing this material for advertising or promotional
purposes, creating new collective works, for resale or redistribution to servers
or lists, or reuse of any copyrighted component of this work in other works.

Supplementary downloadable material is provided by the authors at https://github.com/emilbjornson/twoway-relaying-hardware-impairments. The material includes Matlab code that reproduces all simulation results.

M.~Matthaiou and A.~Papadogiannis are with the Department of Signals and Systems, Chalmers University of Technology, SE-412 96, Gothenburg, Sweden (email: michail.matthaiou@chalmers.se, apapadogiannis@gmail.com).}				
\thanks{E.~Bj\"ornson and M.~Debbah are with the Alcatel-Lucent Chair on Flexible Radio, SUPELEC, Gif-sur-Yvette, France (email: \{emil.bjornson, merouane.debbah\}@supelec.fr). E.~Bj\"ornson is also with ACCESS, Signal Processing Lab, KTH Royal Institute of Technology, Stockholm, Sweden.}
}
\maketitle

\markboth{IEEE COMMUNICATIONS LETTERS, VOL. 17, NO. 6, JUNE 2013}%
{Matthaiou \MakeLowercase{\textit{et al.}}: IEEE COMMUNICATIONS LETTERS}

\begin{abstract}

Hardware impairments in physical transceivers are known to have a deleterious effect on communication systems; however, very few contributions have investigated their impact on relaying. This paper quantifies the impact of transceiver impairments in a two-way amplify-and-forward configuration. More specifically, the effective signal-to-noise-and-distortion ratios at both transmitter nodes are obtained. These are used to deduce exact and asymptotic closed-form expressions for the outage probabilities (OPs), as well as tractable formulations for the symbol error rates (SERs). It is explicitly shown that non-zero lower bounds on the OP and SER exist in the high-power regime---this stands in contrast to the special case of ideal hardware, where the OP and SER go asymptotically to zero.
\end{abstract}

\begin{keywords}
Amplify-and-forward, outage probability, symbol error rate, transceiver impairments, two-way relaying.
\end{keywords}

\vspace{-3pt}
\section{Introduction}

Relays can bring significant performance gains to wireless networks in a cost-efficient manner; for example, coverage extension, spatial diversity gains, and uniform quality-of-service \cite{Yang2009}. In the classic half-duplex mode, the transmission between a source and a destination occupies two time slots, thus the effective system throughput (in bits/channel use) is reduced by a factor of two. Two-way relaying, which allows two nodes to communicate in two time slots with the aid of a relay node, can be used to tackle this problem \cite{Louie2010,Liang2012,Yang2012}. In the first time slot, the two nodes transmit information simultaneously to the relay, and the relay broadcasts the information to the designated destinations in the second time slot.

Most research contributions in the area of relaying assume that the transceiver hardware of the relay node is perfect. However, in practice, the transceiver hardware of wireless nodes is always affected by impairments; for example, IQ imbalance, amplifier amplitude-amplitude non-linearities, and phase noise \cite{Schenk2008a,Studer2010a,Zetterberg2011a}. Impairments create a fundamental capacity ceiling that cannot be crossed by increasing the transmit power; thus, they have a very significant impact especially in high-rate systems \cite{Bjornson2013c}. Since relays are desirable to be low-cost equipment, their transceiver hardware would be of lower quality and, hence, more prone to impairments. Despite the importance of impairments for relaying, there are very few relevant works and these only investigate their impact on one-way relaying. In this context, \cite{Qi2012a,Bjornson2013icassp} (and references therein) analyzed how transceiver impairments affect the symbol error rate (SER) and outage probability (OP), respectively, in one-way relaying.

Motivated by the above discussion, we hereafter analytically assess the impact of relay transceiver impairments in a two-way relaying configuration, considering the amplify-and-forward (AF) protocol. More specifically, we obtain expressions for the signal-to-noise-and-distortion ratio (SNDR) at both transmitter nodes, as well as closed-form expressions for the exact and asymptotic OP/SER. This enables an accurate characterization of the impact of transceiver hardware impairments on both metrics. Our asymptotic analysis provides engineering insights on how the maximal communication performance varies with the level of impairments. To the best of our knowledge, this is the first paper analyzing the impact of hardware impairments in a two-way relaying configuration. The analysis utilizes the generalized impairment model of \cite{Schenk2008a}.

\vspace{-3pt}
\section{Signal and System Model}
\begin{figure} 
\begin{center}
\includegraphics[width=\columnwidth]{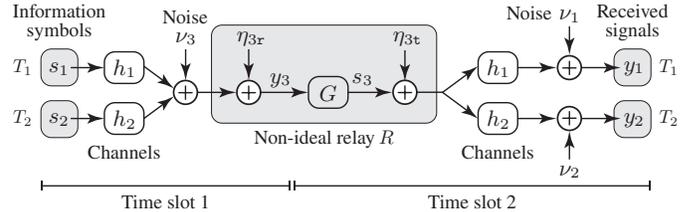}
\end{center} \vskip-3mm
\caption{Block diagram of two-way AF relaying with a non-ideal relay with hardware impairments, represented by the distortion noise terms $\eta_{3\tt{t}},\eta_{3\tt{r}}$.} \label{figure:block-model} \vskip-4mm
\end{figure}

This paper considers two-way AF relaying systems, where two nodes exchange information through a relay as in Fig.~\ref{figure:block-model}.

\vspace{-5pt}
\subsection{Preliminaries on Distortion Noise from Impairments}
\label{subsec:preliminaries_distortion}
\vspace{-3pt}

Suppose an information symbol $s \in \mathbb{C}$ is conveyed over a wireless channel $h \in \mathbb{C}$ with additive noise $\nu \in \mathbb{C}$. In practice, physical radio-frequency (RF) transceivers suffer from impairments that 1) create a mismatch between the intended symbol $s$ and what is actually generated and emitted; and 2) distort the received signal during the reception processing.

Detailed models can be created for each source of impairments (e.g., IQ imbalance, amplifier non-linearities, and phase-noise) in a given hardware setup \cite{Schenk2008a}. However, the combined influence is often well-modeled by a generalized channel model, where the received signal is
\begin{equation}  \label{eq_generalized_model}
y = h (s+\eta_{\tt{t}}) + \eta_{\tt{r}} + \nu
\end{equation}
while $\eta_{\tt{t}},\eta_{\tt{r}}$ are \emph{distortion noises} from impairments in the transmitter and receiver, respectively \cite{Schenk2008a}. Theoretical investigations and measurements (e.g., \cite{Studer2010a,Zetterberg2011a}) have shown that
\begin{equation} \label{eq_model_impairments}
 \eta_{{\tt t }} \sim \mathcal{CN}(0, \kappa_{{\tt t }}^2 P ), \quad \eta_{{\tt r}} \sim \mathcal{CN}(0, \kappa_{{\tt r }}^2 P |h|^2 )
\end{equation}
where $P= \mathbb{E}_s\{ |s|^2 \}$ and the circularly-symmetric complex Gaussianity can be explained by the aggregate effect of many impairments.\footnote{The Gaussianity also holds for the residual distortion when compensation algorithms are applied to handle multiplicative distortion errors \cite{Studer2010a}.} Note that $\mathbb{E}\{\cdot\}$ denotes the expectation operator.

The design parameters $\kappa_{{\tt t }},\kappa_{{\tt r }} \geq 0$ characterize the \emph{level of impairments} in the transmitter and receiver hardware, respectively. These  parameters can be interpreted as the error vector magnitudes (EVMs). EVM is a common quality measure of RF transceivers and defined as the ratio of distortion to signal magnitude.\footnote{The EVM at the transmitter is defined as $\sqrt{\mathbb{E}_{\eta_{\tt{t}}}\{ |\eta_{\tt{t}}|^2 \} / \mathbb{E}_s \{|s|^2\}}$. 3GPP LTE has EVM requirements in the range $\kappa_{\tt{t}} \in [0.08,0.175]$, where smaller values are needed to support the highest spectral efficiencies \cite[Sec.~14.3.4]{Holma2011a}.} Note that \eqref{eq_generalized_model} reduces to the classical model $y = hs + \nu$, when  $\kappa_{{\tt t }}\!=\!\kappa_{{\tt r }}\!=\!0$ (i.e., the case of ideal hardware).

\begin{remark}
The distortions from transceiver impairments act as an additional noise source of variance $(\kappa_{{\tt t }}^2 + \kappa_{{\tt r }}^2 ) P |h|^2$, thus it is sufficient to determine the \emph{aggregate level of impairments} $\kappa = \sqrt{\kappa_{{\tt t }}^2 + \kappa_{{\tt r }}^2}$. We consider hardware impairments only at the relay, for the sake of brevity in interpretation.
The analysis can be readily generalized by reinterpreting the $\kappa$-parameters in our setup as the aggregate parameters for each link.
\end{remark}

\vspace{-3pt}
\subsection{Two-Way Relaying with Ideal and Non-Ideal Hardware}

We consider a two-way AF relaying configuration involving two transmitter nodes ($T_1$ and $T_2$) and a relay node ($R$). Communication takes place in two time slots, where in the first time slot $T_1$ and $T_2$ transmit the information symbols $s_1$ and $s_2$, respectively, to $R$. The relay receives a superimposition of the symbols and broadcasts an amplified version of it to $T_1$ and $T_2$ in the second time slot. A block diagram is given in Fig.~\ref{figure:block-model}.
For brevity, the intended receiver of $s_i$ will be denoted as $T_{r_i}$, where $r_i \triangleq \frac{2}{i}$ for $i=1,2$. We use the subscripts $1$, $2$, $3$ to refer to terms associated with $T_1$, $T_2$, and $R$, respectively.

The signal received at $R$ in the first time slot is given by
\begin{align} \label{relay_rec}
	y_3 = h_1 s_1 + h_2 s_2 + \eta_{3\tt{r}} + \nu_3
\end{align}
where $s_i \sim \mathcal{CN}(0,P_i)$, for $i=1,2$, is an information symbol from a zero-mean circularly-symmetric complex Gaussian distribution with power $P_i$.
In addition, $\nu_i\sim\mathcal{CN}(0,N_i)$ represents the additive complex Gaussian noise at $T_1$, $T_2$, and $R$ for $i=1,2,3$. The channel coefficient for the link $T_i \rightarrow R$ (and for the reciprocal link $R \rightarrow T_i$) is denoted by $h_i$ for $i=1,2$. Each of them is modeled as independent \emph{Rayleigh fading distributed} with average gain $\Omega_i=\mathbb{E}_{h_i} \{|h_i|^2\}$, which means that $h_i \sim \mathcal{CN}(0,\Omega_i)$. As such, the probability density function (PDF) and cumulative density function (CDF) of the channel gains, $\rho_i \triangleq |h_i|^2$, are respectively given by
\begin{align}
f_{\rho_i}(x) = \frac{1}{\Omega_i}e^{-\frac{x}{\Omega_i}},  \quad F_{\rho_i}(x) = 1 - e^{-\frac{x}{\Omega_i}}, \quad x \geq 0. \label{eq_cdf_exp}
\end{align}

Based on the model in Section \ref{subsec:preliminaries_distortion}, the distortion noise in the receiver hardware of the relay is given by the term $\eta_{3\tt{r}} \sim \mathcal{CN}\big( 0, \kappa_{3\tt{r}}^2 (\rho_1 P_1  +  \rho_2 P_2) \big)$ in \eqref{relay_rec}.
In the second time slot, the transmitted signal, $s_3$, by $R$ is simply an amplified version of its received signal $y_3$, or $s_3 = Gy_3$. We assume that all nodes have perfect instantaneous knowledge of the fading channels $h_1, h_2$. Thus, $R$ can apply variable gain relaying,
\begin{align}
	G \triangleq \sqrt{\frac{P_3}{(\rho_1 P_1 + \rho_2P_2)(1+\kappa_{3\tt{r}}^2) + N_3}} \label{eq:relaying_gain}
\end{align}
where $P_3$ is the average transmit power of the relay node. Note that the level of impairments $\kappa_{3\tt{r}}$ in \eqref{eq:relaying_gain} may not be perfectly known.
Such a potential mismatch will degrade the system performance and can be easily incorporated in the subsequent analysis.
We can express the signals received at $T_1$ and $T_2$ as
	\begin{align}
	\!\!  y_{i} &= h_{i} \left(G y_3 + \eta_{3\tt{t}} \right)+\nu_{i} \notag\\
	&= G h_1h_2 s_{r_i} + G h_{i}^2 s_{i} + G h_{i} (\eta_{3\tt{r}} + \nu_3) + h_{i} \eta_{3\tt{t}} +\nu_{i} \label{eq:recieved_signals}
	\end{align}
for $i=1,2$, where $\eta_{3\tt{t}}  \sim \mathcal{CN}(0,\kappa_{3\tt{t}}^2 P_3)$ models distortion noise in the transmitter hardware of the relay. Note that \eqref{eq:recieved_signals}
simplifies to $y_{i} = G h_1h_2 s_{r_i} + G h_{i}^2 s_{i} + G h_{i} \nu_3 +\nu_{i}$ under ideal hardware; this special case was considered in \cite[Eqs.~(2)--(3)]{Louie2010}.

The node $T_{i}$ wants to extract $s_{r_i}$ from $y_{i}$. Since it knows its own transmitted symbol $s_{i}$, it can perfectly eliminate the corresponding self-interference term $G h_{i}^2 s_{i}$. Then, the effective SNDR at $T_{i}$ for detection of the symbol $s_{r_i}$ is
\begin{equation}
{\tt SNDR}_i = \frac{ \rho_1\rho_2 P_{r_i} }{\rho_{i} (N_3 + \kappa_{3\tt{r}}^2 (\rho_1 P_1 + \rho_2P_2) )  + \frac{\rho_{i} \kappa_{3\tt{t}}^2 P_3 + N_{i}}{G^2}}. \label{eq:SNDR_withG}
\end{equation}
By substituting \eqref{eq:relaying_gain} into \eqref{eq:SNDR_withG}, we obtain
\begin{equation}
{\tt SNDR}_i = \frac{ \rho_1\rho_2 }{ \rho_{i}^2 \frac{P_{i}}{P_{r_i}} c \!+\! \rho_1 \rho_2 c \!+\! \rho_{r_i}  b_{i} \!+\! \rho_{i} \left( a_{i} \!+\!  \frac{P_{i}}{P_{r_i}} b_{i} \right)  \!+\! \frac{N_{i} N_3}{P_{r_i} P_3}} \label{eq:SNDR_withoutG}
\end{equation}
where $a_{i} \triangleq \frac{N_3}{P_{r_i}} (1+\kappa_{3\tt{t}}^2)$, $b_{i} \triangleq \frac{N_{i}}{P_3} (1+\kappa_{3\tt{r}}^2)$, and $c \triangleq \kappa_{3\tt{t}}^2 + \kappa_{3\tt{r}}^2 + \kappa_{3\tt{t}}^2 \kappa_{3\tt{r}}^2$ for $i=1,2$. In the special case of ideal hardware, \eqref{eq:SNDR_withoutG} reduces to the expression in \cite[Eqs.~(8)--(9)]{Louie2010}.

\section{Performance Analysis}

\subsection{Exact Outage Probability Analysis}

The OP at $T_{i}$ is denoted by $ {P}_{{\tt out},i}(\thres)$ and is the probability that
channel fading makes ${\tt SNDR}_i$ in \eqref{eq:SNDR_withoutG} fall below a certain threshold, $\thres$, of acceptable communication quality; that is,
\begin{equation} \label{eq:outage_prob_def}
 {P}_{{\tt out},i}(\thres) \triangleq \Pr \{ {\tt SNDR}_i \leq \thres\}.
\end{equation}

\begin{proposition} \label{prop:outage-probability}
The outage probability at $T_{i}$ (when acquiring $s_{r_i}$) is given by $ {P}_{{\tt out},i}(\thres) = 1$ for $\thres \geq \frac{1}{c}$ and
\begin{equation} \label{eq:outage-probability}
\begin{split}
 {P}_{{\tt out},i}(\thres)& = 1 - e^{- \big( \frac{\thres}{1-c\thres} \left(\frac{a_{i}}{\Omega_{r_i}} + \frac{b_{i}}{\Omega_{i}}\right) +\frac{\thres (1+c\thres)}{(1-c\thres)^2} \frac{b_{i}}{\Omega_{r_i}} \frac{P_{i}}{P_{r_i}} \big)} \\
& \times 2 \sqrt{ \frac{\frac{(\thres+\thres^2)}{(1-c\thres)^2} \frac{N_{i} N_3}{\Omega_1 \Omega_2 P_{r_i} P_3}  + \frac{\thres^2}{(1-c\thres)^3} \frac{b_{i}^2 P_{i}}{\Omega_1 \Omega_2 P_{r_i}} } { 1 +  \frac{c\thres}{(1-c\thres)} \frac{P_{i} \Omega_{i}}{P_{r_i} \Omega_{r_i}}} }   \\ &\times K_1 \left( 2 \sqrt{ \frac{\frac{(\thres+\thres^2)}{(1-c\thres)^2} \frac{N_{i} N_3}{\Omega_1 \Omega_2 P_{r_i} P_3}  + \frac{\thres^2}{(1-c\thres)^3} \frac{b_{i}^2 P_{i}}{\Omega_1 \Omega_2 P_{r_i}} } { \Big( 1 +  \frac{c\thres}{(1-c\thres)} \frac{P_{i} \Omega_{i}}{P_{r_i} \Omega_{r_i}} \Big)^{-1} } } \right)
\end{split}
\end{equation}
for $0 \leq \thres < \frac{1}{c}$, where $K_1(\cdot)$ denotes the first-order modified Bessel function of the second kind.
\end{proposition}
\begin{IEEEproof}
Referring to \eqref{eq:outage_prob_def}, we see that $ {P}_{{\tt out},i}(\thres)$ depends on ${\tt SNDR}_i$, which is a function of the two independent random variables $\rho_1$ and $\rho_2$. Using the law of total probability to condition on $\rho_{i}$, we obtain the expression $\Pr \{ {\tt SNDR}_i \leq \thres \} = 1 - \int_{0}^{\infty} \Pr \{ {\tt SNDR}_i > \thres | \rho_{i} \} f_{\rho_{i}}(\rho_{i}) d \rho_{i}$.

To evaluate this integral, note that $\Pr \{ {\tt SNDR}_i > \thres | \rho_{i} \} =$
\begin{equation}
\! \begin{cases} 1\!-\! F_{\rho_{r_i}} \!\bigg( \!\frac{\thres \big( \rho_{i}^2 \frac{P_{i}}{P_{r_i}} c + \rho_{i} \big( a_{i} + \! \frac{P_{i}}{P_{r_i}} b_{i} \big)  + \frac{N_{i} N_3}{P_{r_i} P_3} \big) }{ \rho_{i} (1-c\thres)-b_{i}} \!\bigg), & \!\! \text{if }\thres < \frac{1}{c},  \!\!\! \\
0, & \!\! \text{if } \thres \geq \frac{1}{c}, \!\!\!
\end{cases}
\end{equation}
after some algebra. Using the PDF and CDF in \eqref{eq_cdf_exp}, we obtain \eqref{eq:outage-probability} by making a change of variables $z = \rho_{i} - \frac{b_{i}}{(1-c\thres)}$ and then evaluating the integral using \cite[Eq.~(3.324.1)]{Gradshteyn2007a}.
\end{IEEEproof}

Proposition \ref{prop:outage-probability} provides a new and tractable closed-form OP expression under the presence of transceiver hardware impairments at the relay. It is a generalization of \cite[Theorem 1]{Louie2010}, where the special case of ideal hardware was considered.

\vspace{-8pt}
\subsection{Asymptotic Outage Probability Analysis} \label{subsection:asymptotic-OP}
In order to obtain some engineering insights into the fundamental impact of impairments, we now elaborate
on the high-power regime. In this case, we assume, without significant loss of generality, that
$P_1=P_2=\tau P_3$ grow large (with $\tau>0$), which means that the relaying gain $G$ in \eqref{eq:relaying_gain} converges to
\begin{align}
	G^{\infty} = \sqrt{\frac{1}{\tau(\rho_1+\rho_2)(1+\kappa_{3\tt{r}}^2) }}
\end{align}
and remains finite and strictly positive. It is easy to see that the SNDR in \eqref{eq:SNDR_withoutG} becomes asymptotically equal to
\begin{equation} \label{eq:SNDR_withoutG_asymptotic}
{\tt SNDR}^{\infty}_i = \frac{ \rho_1\rho_2 }{ \rho_{i}^2 c + \rho_1 \rho_2 c} = \frac{\rho_{r_i}}{ (\rho_1 + \rho_2) c}.
\end{equation}

\begin{corollary} \label{cor:outage-high-power}
In the high-power regime where $P_1=P_2=\tau P_3 \rightarrow \infty$ and $\tau >0$, the outage probability at $T_{i}$ becomes
\begin{equation} \label{eq_pout_as_fg}
 {P}_{{\tt out},i}^\infty (\thres) =  \begin{cases} \frac{\Omega_{i} c \thres }{\Omega_{r_i} + c \thres (\Omega_{i}-\Omega_{r_i})}, &  \text{if } \thres<\frac{1}{c},\\
1, &   \text{if } \thres \geq \frac{1}{c}. \end{cases}
\end{equation}
\end{corollary}
\begin{IEEEproof}
Similar as for Proposition \ref{prop:outage-probability}, but using \eqref{eq:SNDR_withoutG_asymptotic}.
\end{IEEEproof}

Two important observations can be made from Corollary~\ref{cor:outage-high-power}. Firstly, $ {P}_{{\tt out},i}(x)=1$ for $\thres \geq \frac{1}{c}$, thus there is a (finite) upper bound on the SNDR, \emph{an SNDR ceiling}, in the high-power regime. The value of this upper bound is characterized by the level of impairments, because $c \triangleq \kappa_{3\tt{t}}^2 + \kappa_{3\tt{r}}^2 + \kappa_{3\tt{t}}^2 \kappa_{3\tt{r}}^2$. Note that this fundamental phenomenon disappears in the special case of ideal hardware, since $\kappa_{3\tt{t}}=\kappa_{3\tt{r}}=0$ makes $c=0$.

Secondly, there is a non-zero lower bound on the OP for $0 \leq \thres < \frac{1}{c}$. The value of this bound depends on the level of impairments, the average channel gains $\Omega_1,\Omega_2$, and the threshold $\thres$. This is unique for two-way relaying and is due to an unresolvable ambiguity created by transceiver impairments; more precisely, the presence of impairments creates two distortion noises with different variances at the relay node (one per transmitter). These are amplified by the second hop, such that the randomness does not vanish asymptotically, as seen from the presence of $\rho_1,\rho_2$ in \eqref{eq:SNDR_withoutG_asymptotic}. The remaining randomness implies that the OP will not converge to zero. In contrast, the OP does approach zero in the high-power regime for both two-way relaying with ideal hardware \cite{Louie2010} and one-way relaying with transceiver impairments \cite{Bjornson2013icassp}, since the asymptotic SNDRs become deterministic in these scenarios.

From the perspective of hardware design, suppose we have a total EVM constraint $\kappa_{\tt{tot}}>0$ on the relay, such that $\kappa_{3\tt{t}} + \kappa_{3\tt{r}}=\kappa_{\tt{tot}}$.
Corollary 3 in \cite{Bjornson2013icassp} proves that $c$ is minimized by $\kappa_{3\tt{t}} = \kappa_{3\tt{r}} = \frac{\kappa_{\tt{tot}}}{2}$; thus, the asymptotic OP is minimized by selecting transmitter/receiver hardware of the same quality.

\begin{remark}
Note that practical systems suffer from other non-hardware related impairments as well,
such as channel estimation errors \cite{Wang2012}. Such additional impairments will only degrade
the performance, thus the results herein serve as an upper
bound on what is achievable in practice.
\end{remark}

\subsection{Exact and Asymptotic Symbol Error Rate Analysis}

We now turn our attention to the SER. To this end, we first invoke that for many modulation formats (e.g., BPSK, BFSK with orthogonal signaling, and $M$-ary PAM),
the average SER at $T_i$ can be represented by the generic formula
\begin{equation}
{\tt SER}_i=\mathbb{E}_{{\tt SNDR}_i} \! \left\{\alpha Q\left(\sqrt{2\beta{\tt SNDR}_i}\right)\right\}, \quad i=1,2
 \label{eq:SER_general}
\end{equation}
where $Q(x)=\frac{1}{\sqrt{2\pi}}\int_{x}^\infty e^{-t^2/2}dt$ is the Gaussian $Q$-function and $\alpha, \beta$ are modulation specific constants. Using integration by parts,  \eqref{eq:SER_general} can be reformulated into the mathematically more convenient form
\begin{align}\label{eq_SER_alt}
	{\tt SER}_i=\frac{\alpha\sqrt{\beta}}{2\sqrt{\pi}}\int_0^\infty \frac{e^{-\beta x}}{\sqrt{x}} F_{{\tt SNDR}_i}(x) dx
\end{align}
where the CDF of ${\tt SNDR}_i$ is $F_{{\tt SNDR}_i}(x) =  {P}_{{\tt out},i}(x)$ by definition.
Combining Proposition 1 and \eqref{eq_SER_alt}, it does not appear that the resulting integral can be evaluated in closed-form; however, the SER
can be obtained from \eqref{eq_SER_alt} by numerical integration which is much more efficient than Monte-Carlo simulations.

We henceforth consider the high-power regime (as defined in Section \ref{subsection:asymptotic-OP}) and obtain the following result.

\begin{corollary} \label{cor:SER-high-power}
Consider the high-power regime where $P_1=P_2=\tau P_3 \rightarrow \infty$. For identical average channel gains $\Omega_1=\Omega_2$, the SERs at $T_1$ and $T_2$ are identical and equal to \vskip-4mm
\begin{equation} \label{eq:asymptotic-SER}
{\tt SER}_1^\infty = {\tt SER}_2^\infty = \frac{\alpha c}{2 \beta \sqrt{\pi}} \gamma\left(\frac{3}{2},\frac{\beta}{c}\right) + \frac{\alpha}{2} \textrm{erfc}\left( \! \sqrt{\frac{\beta}{c}} \right)
\end{equation}
where $\gamma(p,x)=\int_0^{x} t^{p-1}e^{-t}dt$ is the lower incomplete gamma function and
$\textrm{erfc}(x)=\frac{2}{\sqrt{\pi}}\int_{x}^\infty e^{-t^2}dt$ denotes the complementary error function.
\begin{IEEEproof}
The asymptotic SER in \eqref{eq:asymptotic-SER} follows from computing \eqref{eq_SER_alt} using the OP in \eqref{eq_pout_as_fg} for $\Omega_1=\Omega_2$.
\end{IEEEproof}
\end{corollary}

The asymptotic SERs in Corollary \ref{cor:SER-high-power} are strictly positive, except for the special case of ideal hardware ($c=0$). Hence, hardware impairments create an irreducible error floor.

Suppose we want to create a practical two-way relaying system that supports a given set of OP and SER requirements. By inverting the asymptotic expressions in Corollaries \ref{cor:outage-high-power} and \ref{cor:SER-high-power} with respect to $c$, we can determine fundamental design guidelines on the \emph{highest} level of impairments (in terms of $c$) that can theoretically meet the stipulated requirements.

\section{Numerical Results}\label{section:numerical-results}

\begin{figure}[!t] \vskip-1mm
\begin{center}
\includegraphics[width=\columnwidth]{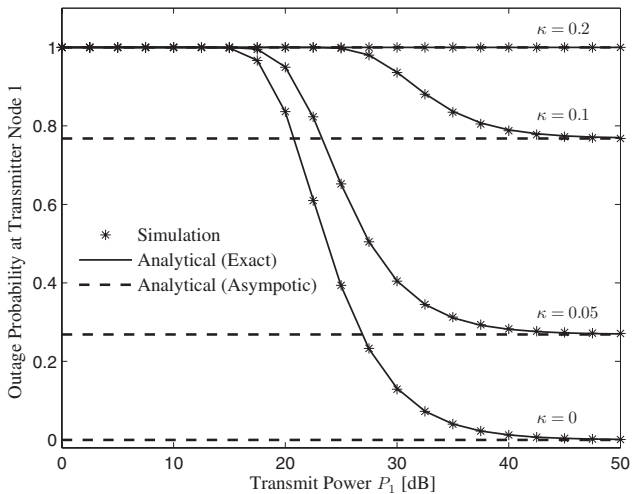}
\end{center} \vskip-6mm
\caption{Outage probability (OP) at node $T_1$ against the transmit power $P_1$. Simulation parameters: $x = 2^5-1$, $\Omega_1=2$, $\Omega_2=1$, and $P_1=P_2=2P_3$.}
\label{figure:OP1} \vskip-3mm
\end{figure}

We now present a set of numerical results to validate our previous theoretical results.
We consider a scenario with symmetric signal and noise powers: $P_1=P_2=2P_3$ and $N_i=1$ for $i=1,2,3$.
Fig.~\ref{figure:OP1} compares the simulated OP at $T_1$ against the exact expression of Proposition \ref{prop:outage-probability} and the asymptotic limit of Corollary \ref{cor:outage-high-power}. We consider a high-rate system with $x=2^5-1$ (i.e., 5 bits/channel use) and different levels of impairments $\kappa_{3\tt{t}}= \kappa_{3\tt{r}}=\kappa$. We can observe how dramatic the impact of impairments can be in high-rate systems. With a high level of impairments of $\kappa=0.2$, the system is always in outage and no communication can be supported---irrespective of the transmit power level. At moderate level of impairments, the OP approaches a non-zero saturation value in the high-power regime, that is accurately predicted by Corollary \ref{cor:outage-high-power}. This stands in fundamental contrast with the case of ideal hardware $\kappa=0$, where the OP goes asymptotically to zero \cite{Louie2010}.

\begin{figure}[!ht] \vskip-1mm
\begin{center}
\includegraphics[width=\columnwidth]{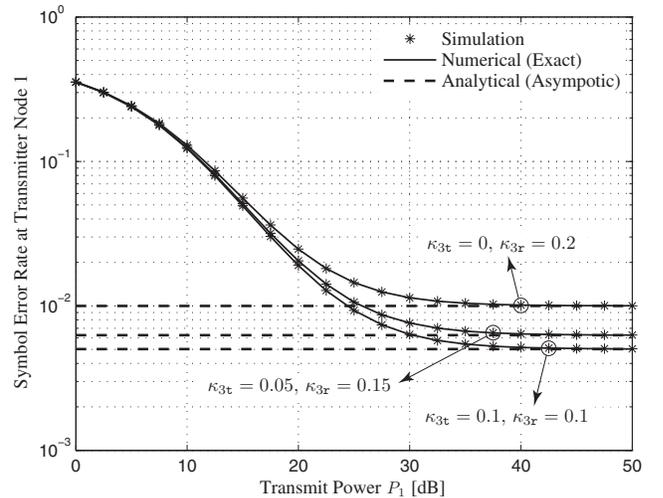}
\end{center} \vskip-6mm
\caption{Symbol error rate (SER) at node $T_1$ against the transmit power $P_1$. Simulation parameters: BPSK modulation, $\Omega_1\!=\!\Omega_2\!=\!1$, and $P_1\!=\!P_2\!=\!2P_3$.}
\label{figure:SER1} \vskip-3mm
\end{figure}

In Fig.~\ref{figure:SER1}, we consider the SER with BPSK modulation (i.e., $\alpha=\beta=1$) and investigate different impairment combinations for which $\kappa_{3\tt{t}} + \kappa_{3\tt{r}}=0.2$ is constant. The exact curves are obtained by numerical evaluation of \eqref{eq_SER_alt}, while the high-power SER limits stem from Corollary \ref{cor:SER-high-power}. As anticipated, the best choice for minimizing the SER is to have the same hardware quality ($\kappa_{3\tt{t}} \!=\! \kappa_{3\tt{r}} \!=\! 0.1$) at the transmit and receive side of the relay. Such a choice asymptotically reduces the SER by a factor of 2, compared to the case where $\kappa_{3\tt{t}}=0, \kappa_{3\tt{r}}=0.2$. Generally, a relay node with low-quality hardware on one side and high-quality hardware on the other side should be avoided.

\section{Conclusions}
This paper investigated analytically the impact of transceiver hardware impairments on two-way relaying systems. Closed-form and tractable expressions for the OP and SER were obtained as a function of the level of impairments. Our theoretical analysis indicated that the receive and transmit hardware of the relay node should be of the same quality to minimize both these metrics. It was also shown that in high-rate systems and/or with high level of impairments, the system is always in full outage. Moreover, zero OP or SER cannot generally be achieved, since the SNDR is random even in the high-power regime. Ultimately, our simple asymptotic expressions can provide engineering guidelines to select hardware that satisfies the OP/SER requirements of a practical relaying topology.

\section*{ACKNOWLEDGMENTS}
E. Bj\"ornson is funded by the International Postdoc Grant 2012-228 from the Swedish Research Council.

\bibliographystyle{IEEEbib}
\bibliography{IEEEabrv,refs}
\end{document}